\documentstyle[prl,aps]{revtex}      

\begin{document}

\draft  
\tighten
\twocolumn[\hsize\textwidth\columnwidth\hsize\csname@twocolumnfalse%
\endcsname

\title{Probing Microquasars with TeV Neutrinos}

\author{Amir Levinson$^1$ \& Eli Waxman$^2$}

\address{$^1$School of Physics and Astronomy, Tel Aviv University,
Tel Aviv 69978, Israel\hfil}
\address{$^2$Department of Condensed Matter Physics, Weizmann Institute, 
Rehovot 76100, Israel}

\maketitle 

\begin{abstract}

The jets associated with Galactic micro-quasars are believed to be 
ejected by accreting stellar mass black-holes or neutron stars. 
We show that if the energy content of the jets in the transient sources is
dominated by electron-proton plasma, then a several hour 
outburst of 1--100~TeV neutrinos produced by photo-meson interactions
should precede the radio flares associated with major ejection events.
Several neutrinos may be detected during
a single outburst by a 1~km$^2$ detector, thereby 
providing a powerful probe of
micro-quasars jet physics. 

\end{abstract}

]

\narrowtext

Microquasars are Galactic jet sources associated with some classes of 
X-ray binaries involving both, neutron stars and black hole candidates 
(albeit with some notable differences between these 
two classes)\cite{MR99,FK01,Fe01}.
During active states, the X-ray flux and spectrum can vary substantially
between different sub-states, with a total luminosity that, during the 
so-called very high states, often exceeds the Eddington limit 
(typically a few times $10^{38}$ ergs s$^{-1}$ in those sources).
Their activity involves ejection of jets with kinetic 
power that appears to constitute a considerable fraction of the liberated
accretion energy (in some cases the minimum jet power inferred exceeds
the peak X-ray luminosity \cite{LB96,AA99}), and 
that gives rise to intense radio and IR flares.  Radio monitoring of some
X-ray transients have revealed superluminal motions in currently 
three of the sources \cite{MR94,HR95,Hetal00}, indicating that at least in 
these objects the jets are relativistic.  The Lorentz factors of the radio 
emitting blobs have been estimated to be (coincidently) $\Gamma\sim 2.5$ 
in the two superluminal sources, GRS 1915+105 
\cite{MR94} and GRO J1655-40 
\cite{HR95}, and somewhat smaller ($\Gamma\sim 2$) in the third one, XTE J1819
\cite{Hetal00}.  Whether these are 
representative values or merely the result of selection effects is unclear
at present; the class of microquasars may contain sources with much larger
$\Gamma$, rendered invisible by beaming away from us \cite{LB96}.  

The temporal behavior of microquasars appears to be rather complex.  They
exhibit large amplitude variations over a broad range of time scales
and frequencies, with apparent connections between the radio, IR, and 
soft/hard X-ray fluxes \cite{Hetal97,Eetal98,Fetal99,Ya01}.  The characteristics of 
the multi-waveband behavior depend on the state of the source, that is, whether 
the source is in a very high, soft/high or low/hard state \cite{Fe01}.  The 
ejection episodes are classified into several classes according to the brightness 
of synchrotron emission produced in the jet and the characteristic time scale 
of the event \cite{Eetal00}.  The duration of 
major ejection events (class A) is typically on the order of days, while
that of less powerful flares (classes B and C) is correspondingly shorter
(minutes to hours).
The correlations between the X-ray and synchrotron emission clearly indicates 
a connection between the accretion process and the jet activity.  Whether 
radio and IR outbursts represent actual ejection of blobs of plasma 
or, alternatively, formation of internal shocks in a quasi-steady jet is 
unclear (cf. ref. \cite{KSS00}).  In any case, since the overall time scale 
of outbursts (minutes to days) 
is much longer than the dynamical time of the compact object (milliseconds),
it is likely that shocks will continuously form during the ejection event
over a range of time scales that encompass the dynamical time,
owing to fluctuations in the parameters of the expelled wind, leading to 
dissipation of a substantial fraction of the bulk energy at relatively 
small radii.  The extremely rapid variations of the X-ray flux
often seen in these sources supports this view (e.g., ref \cite{NARetal01}).  
If a fraction of at least a 
few percent of the jet power is tapped for acceleration of electrons to 
very high energies, then emission of high-energy gamma-rays is anticipated,
in addition to the observed radio and IR emission \cite{LB96,AA99}.  EGRET upper
limits for some X-ray novae have been considered in this context \cite{LM96}. 
A recently discovered microquasar which appears to have a persistent radio 
jet \cite{PMRM00}, seems to coincide with an unidentified EGRET source having 
a total luminosity in excess of its X-ray luminosity. 

The content of jets in microquasars is yet an open issue.  The 
synchrotron emission both in the radio and in the IR is consistent with
near equipartition magnetic field, which is also implied by minimum energy
considerations \cite{LB96}.  However, the dominant energy carrier in the jet is 
presently unknown (with the exception of the jet in SS433).  Scenarios 
whereby energy extraction is associated with spin down of a Kerr black 
hole favor $e^{\pm}$ composition (although baryon entrainment is an issue).  
However, the pair annihilation rate inferred from the estimated jet power
implies electromagnetic domination on scales smaller than roughly 10$^9$
cm in the superluminal sources, and requires a transition from
electromagnetic to particle dominated flow above the annihilation radius
by some unspecified mechanism \cite{LB96}.  Alternatively, in scenarios in which an 
initial rise of the x-ray flux leads to ejection of the inner part of the accretion 
disk, as widely claimed to be suggested by the anti-correlation between 
the X-ray and radio flares seen during major ejection events (cf. ref. \cite{LB96}
for a different interpretation), e-p jets are 
expected to be produced.  A possible diagnostic of e-p jets is the presence 
of Doppler-shifted spectral lines, such as the H$_{\alpha}$ line as seen 
in SS433.  The detection of such
lines from jets having Lorentz factor well in excess of unity (as is the 
case in the superluminal microquasars) may, however, be far more difficult
than in SS433, as the lines are anticipated to be very broad 
($\Delta \lambda/\lambda\gtrsim$ 0.1).  Furthermore, 
the conditions required to produce detectable flux in such sources may be 
far more extreme than in SS433 \cite{LB96b}.  Here we propose another 
diagnostic of 
hadronic jets, namely emission of TeV neutrinos.  As shown below, for 
typical parameters the neutrinos are produced on scales much smaller than
the IR and radio emission and, therefore, can provide a probe of the 
innermost structure of microquasars jets.  

The picture envisaged is the following:  We suppose that on sufficiently 
small scales ($\lesssim 10^{11}$ cm) a significant fraction of the 
energy liberated during ejection events dissipates, e.g. through the
formation of internal collisionless shocks, leading to the acceleration of 
a non-thermal power law
distribution of protons (and electrons) up to the maximum energy achievable.
Detailed studies \cite{BE87} suggest that shock acceleration of protons may 
be highly efficient, and should produce a power law spectrum, 
$dn_p(\epsilon)/d\epsilon\propto \epsilon^{-2}$ (perhaps slightly steeper in the 
case of ultra-relativistic shocks).  The fraction of energy carried by the power 
law component of electrons may depend sensitively on the injection process,
but generally thought to constitute a few percent.  In the following, we denote
by $\eta_P$ the fraction of the total burst energy that is tapped for the 
acceleration of protons to nonthermal energies.  We emphasize that $\eta_p$ 
represents essentially the product of the fraction of total burst energy that 
dissipates behind the shocks (which depends primarily on the duty cycle of shock 
formation) and the efficiency at which protons are accelerated to nonthermal 
energies by a single shock (that may 
approach 100\%). 
Observations of $\gamma$-ray bursts, which are believed
to arise from a similar process of jet kinetic energy dissipation (albeit
with different Lorentz factors and luminosities \cite{Meszaros-Science}), 
suggest that a significant
fraction of the jet energy is dissipated and converted to a power-law
distribution of accelerated electrons \cite{Freedman01}.

As shown below, synchrotron emission by electrons 
contributes large opacity to photo-meson production over a range of radii
that depends on the strength of the magnetic field.  (Since the photo-meson opacity
is contributed predominantly by thermal electrons, the analysis outlined below 
is rendered independent of the details of electron injection to 
nonthermal energies.)  Between 12\% to 25\% of the injected proton 
energy is converted, by virtue of the large optical depth, into muon 
neutrinos having a flat spectrum in the range between 1 and $\sim 100$ 
TeV, which then freely escape the system. 

Consider a jet of kinetic power $L_{j}=10^{38}L_{j38}$ erg s$^{-1}$,
and opening angle $\theta=0.1\theta_{-1}$, with a corresponding proper energy 
density $U_{j}=L_j/(\pi\theta^2 r^2 \Gamma^2 c)$, propagating in the background
of X-ray photons emitted by the accretion disc, with luminosity
$L_x=10^{38}L_{x38}$ erg s$^{-1}$. 
In the scenario under consideration, significant fraction of the jet kinetic
energy is dissipated via mildly-relativistic 
internal shocks in the jet at radii
\begin{equation}
r\gtrsim\Gamma^2 c\delta t=3\times10^7\Gamma^2\delta t_{-3}{\rm\,cm},
\label{eq:rd}
\end{equation}
where $\delta_t=10^{-3}\delta t_{-3}$~s is the source dynamical time.


The Thomson optical depth across the jet is given by
\begin{equation}
\tau_T = \sigma_Tn_pr\theta\Gamma\simeq 0.05 \frac{L_{j38}}{\Gamma r_{8}
\theta_{-1}},
\end{equation}
where  $r=10^8r_8$~cm and
the proper proton number density in the jet is $n_p\simeq U_j/(m_pc^2)$.
External photons can therefore penetrate the jet and interact
with accelerated protons at radii larger than roughly $10^7$ cm. 
The proton energy, in the jet frame, for which interaction with
external X-ray photons is at the $\Delta$-resonance is
\begin{equation}
\epsilon_{p,\Delta}=0.3{\rm GeV}^2/\Gamma\epsilon_x = 3\times10^{14}
(\Gamma\epsilon_x/1{\rm keV})^{-1}{\rm eV},
\end{equation}
where $\epsilon_x$ is the observed photon energy.

The time available, in the jet frame, for photo-pion interactions is
$\approx r/\Gamma c$. Thus, 
the optical depth to photo-pion production at the $\Delta$ resonance, 
contributed by the external photons, is
\begin{equation}
\tau_{p\gamma}\simeq c\sigma_{\rm peak} n_x \frac{r}{\Gamma c}= 
1  \frac{\Gamma L_{x38}} {r_{8}}(\epsilon_x/1{\rm keV})^{-1}, 
\end{equation}
where
$\sigma_{\rm peak}=5\times10^{-28}$ is the cross section at
the $\Delta$-resonance and $n_x$ is the number density of background X-ray 
photons measured in the jet frame:
$n_x\simeq\Gamma(L_x/\epsilon_x)/(4\pi r^2 c)$.
Thus, photo-meson interactions with external X-ray photons at the inner-most 
dissipation radii will convert a large fraction of the accelerated proton 
energy to high energy pions.

Additional contribution to the photo-meson optical depth
comes from synchrotron photons produced inside the
jet by the thermal electrons.  For a relativistic shock, the mean 
electron energy assuming rapid equilibration time is $\sim 0.5
(\Gamma_s-1)m_pc^2$, where $\Gamma_s$ is the Lorentz factor of the shock.  
For the microquasars $\Gamma_s-1\sim 1$ typically.  This then yields 
for the (jet frame) peak energy of the synchrotron spectrum,
\begin{equation}
\epsilon_{\gamma,\rm peak}
\simeq 50\frac{\xi_{-1}^{1/2}L_{j38}^{1/2}}{\Gamma r_{8}
\theta_{-1}}\ \ {\rm keV},
\end{equation}
where $\xi=10^{-1}\xi_{-1}$ is the equipartition parameter, defined through 
the relation $B=\sqrt{8\pi\xi U_j}$.  Note, that $\epsilon_{\gamma,\rm peak}$ 
reduces to infrared energies at a radius
of $\sim 10^{13}$ cm, with a corresponding light crossing time of order 
minutes, compatible with the variability time of IR baby flares.

Assuming the total energy density of synchrotron photons to be on the order 
of $U_j$ yields for the corresponding energy-dependent number density, defined
as $n_{\rm syn.}(\epsilon)=\epsilon (dn_{\rm syn.}/d\epsilon)$,
\begin{eqnarray}
n_{\rm syn.}(\epsilon)\simeq&&\frac{U_j}{\epsilon_{\gamma,\rm peak}}
\left(\frac{\epsilon}{\epsilon_{\gamma,\rm peak}}\right)^{-\alpha}
\cr
\simeq&& 10^{20}\left(\frac{\epsilon}{\epsilon_{\gamma,\rm peak}}
\right)^{-\alpha}
\frac{L_{j38}^{1/2}}{r_8\Gamma\theta_{-1}\xi_{-1}^{1/2}} \ \ {\rm cm^{-3}},
\end{eqnarray}
where $\alpha=1/2$ for $\epsilon<\epsilon_{peak}$, and 
(assuming efficient electron injection) $\alpha=1$ for $\epsilon>\epsilon_{peak}$. 
The spectrum at $\epsilon<\epsilon_{peak}$
results from the fast cooling of electrons. The ratio of electron cooling
time, $t_{\rm syn}\approx 6\pi m_e^2 c/\sigma_T B^2 m_p$, to the dynamical 
time, $r/\Gamma c$, is
\begin{equation}
\frac{t_{\rm syn.}}{r/\Gamma c}
\simeq 10^{-5}\frac{\Gamma^3 r_8\theta_{-1}^2}{L_{j38}\xi_{-1}}.
\end{equation}
We note that at radii $r\lesssim10^9$~cm, inverse-Compton scattering of
electrons is suppressed since scattering is in the Klein-Nishsina regime
for photons carrying significant fraction of the synchrotron energy density. 
At these radii, the fast cooling of electrons indeed implies that the 
energy density of synchrotron photons is of order $U_j$. 
At larger radii, the energy density of synchrotron photons will be 
somewhat suppressed due to inverse-Compton emission.

A rough estimate of the energy loss rate of protons due to photo-meson 
interactions gives: $t_{p\gamma}^{-1}\sim \sigma_{\rm peak}n_{\rm syn.}
c(\Delta\epsilon_p/\epsilon_p)$, where 
$\Delta\epsilon_p/\epsilon_p\simeq0.2$ is 
the average fractional energy loss in a single collision.  
Equating the latter with the acceleration rate $\sim eBc/\epsilon_p$, 
we obtain for the maximum proton energy,
\begin{equation}
\epsilon_{p,\rm max}\simeq5\times10^{15}\xi_{-1}^{1/2}(\Gamma r_8 
\theta_{-1})^{1/3}L_{j38}^{-1/6}\ \ \ {\rm eV}.
\label{eq:Ep-max}
\end{equation}
We emphasize that the proton energy cannot in any case exceed the upper limit 
imposed by the requirement that the protons be confined to the system; viz., 
$\epsilon_p<\theta eBr\simeq2\times10^{16}\xi_{-1}^{1/2}L_{j38}^{1/2}
\Gamma^{-1}$ eV.  Comparing the latter 
with eq. (\ref{eq:Ep-max}) then implies that the maximum 
proton energy is limited by photo-meson losses at radii 
$r_8<10^2 L_{j38}^2\Gamma^{-4}\theta_{-1}^{-1}$, 
and by confinement at larger radii.

The comoving proton energy for which interaction with synchrotron peak 
photons is at the $\Delta$-resonance is
\begin{equation}
\epsilon_{p,\rm peak}= 10^{13}
\frac{\Gamma r_8 \theta_{-1}}{(\xi_{-1}L_{38})^{1/2}}\,{\rm eV}.
\label{eq:Ep-peak}
\end{equation}
Consequently,
protons may be accelerated to energy exceeding that required for photo-meson
interaction with synchrotron peak photons for radii 
\begin{equation}
r\lesssim r_c=5\times10^{11} 
\frac{L_{j38}^{1/2}\xi_{-1}}{\Gamma\theta_{-1}}
\min\left(L_{j38}/\Gamma,\xi_{-1}^{1/2}\right)
\ {\rm cm}.
\label{eq:r_c}
\end{equation}
From eqs. (\ref{eq:Ep-max}) and (\ref{eq:r_c}) it is seen that for our 
choice of parameters protons can be accelerated to energies 
$\sim10^{16}\xi_{-1}$~eV 
in the region where photo-meson interactions take place.  
The contribution of synchrotron photons to the photo-meson optical depth is
\begin{equation}
\tau_{p\gamma}\simeq c\sigma_{\rm peak} n_{\rm syn.} \frac{r}{\Gamma c} 
\simeq \frac{10}{\Gamma^{2}\theta_{-1}}
\left(\frac{L_{j38}}{\xi_{-1}}\right)^{1/2}
\left(\frac{\epsilon_{p}}{\epsilon_{p,\rm peak}}\right)^{\beta}, 
\label{eq:tau-pgamma}
\end{equation}
where $\beta=1/2$ for $\epsilon_{p}>\epsilon_{p,\rm peak}$ and
$\beta=1$ for $\epsilon_{p}<\epsilon_{p,\rm peak}$.

Eq. (\ref{eq:tau-pgamma}) implies that at all radii 
the photo-meson optical depth 
is large for protons of energy exceeding the threshold energy 
(\ref{eq:Ep-peak}) for interaction with synchrotron peak photons.
We therefore expect most of the energy of these protons to be lost
to pion production. Neutral pions and charged pions are produced in the
photo-meson interaction with roughly equal probabilities, and the decay
of a charged pion produces 3 neutrinos, $\pi^+\rightarrow\mu^++\nu_\mu
\rightarrow e^++\nu_e+\overline\nu_\mu+\nu_\mu$,
each carrying $\sim5\%$ of the initial proton energy. Due to the high
photo-meson optical depth, neutrons produced in photo-pion interactions
will lose their energy by photo-meson interactions on a time scale shorter
then the neutron life time. This implies similar production rates of both
negatively and positively charged pions and, hence, similar production rates
of $\nu_\mu$ and $\bar\nu_\mu$.

High energy pions and muons may loose significant fraction of their energy
by inverse-Compton scattering prior to decaying.
Inverse-Compton scattering of synchrotron photons by high energy pions
is in the Klein-Nishina regime for 
\begin{equation}
\epsilon_\pi\gtrsim\epsilon_{\pi,c}=
\frac{(m_\pi c^2)^2}{\epsilon_{\gamma,\rm peak}}\simeq
0.4\frac{\Gamma r_8\theta_{-1}}{(L_{j38}\xi_{-1})^{1/2}}{\rm\,TeV}.
\label{eq:epic}
\end{equation}
Comparing this equation with Eq. (\ref{eq:Ep-peak}), and noting that the
energy of a pion produced by photo-pion interaction is $\approx20\%$ 
of the initial proton energy, we find that pions
produced by protons of energy exceeding the threshold energy satisfy
$\epsilon_\pi\gtrsim\epsilon_{\pi,c}$. 
Thus, the pion inverse-Compton scattering energy loss time is
\begin{equation}
\tau_{IC}\approx\frac{3}{8c\sigma_T n_{\rm syn.}}
\frac{\epsilon_\pi\epsilon_{\gamma,\rm peak}}{(m_e c^2)^2}=
3\times10^{-2}\xi_{-1}\left(\frac{\epsilon_\pi}{1{\rm TeV}}\right){\rm\,s},
\label{eq:tpiIC}
\end{equation}
and the ratio of the pion decay time to energy loss time is
\begin{equation}
\frac{\tau_{IC}}{\tau_{\rm decay}}\approx10^2\xi_{-1}.
\label{eq:tpi}
\end{equation}
We therefore expect pions to decay
prior to significant energy loss. However, muons, the life time of which is 
$\approx100$ times longer, 
may lose significant fraction of their energy before decaying. In the
following we therefore conservatively assume that 
a single high energy $\nu_\mu$ (or $\bar\nu_\mu$) is
produced in a single photo-pion interaction of a proton (or neutron),
corresponding to conversion of $1/8$ of the energy lost to pion
production to muon neutrinos.  In terms of $\eta_p$, the fraction of
jet energy injected as a power law distribution of 
protons, the flux at Earth of $\nu_\mu$ and $\bar\nu_\mu$ can be 
expressed as,
\begin{eqnarray}
{\cal F}_{\nu_\mu}\simeq&&\frac{1}{2}\eta_p\Gamma^{-1}\delta^3\frac{L_j/8}{4\pi D^2}
\cr
=&&0.5\times10^{-9}\eta_{p,-1}\Gamma^{-1}\delta^3 D_{22}^{-2}L_{j38}\ \ 
{\rm erg\ s^{-1} cm^{-2}},
\end{eqnarray}
where $\delta$ is the Doppler factor, $D=10^{22}D_{22}$~cm is the distance 
to the source, and $\eta_p=0.1\eta_{p,-1}$. 
The factor $1/2$ is due to the fact that for a flat
power-law proton distribution, and for the characteristic parameters
invoked, roughly half the energy of the power-law
component is carried by protons which lose most of their energy to 
pion-production [see Eqs. (\ref{eq:Ep-peak},\ref{eq:tau-pgamma})].

Due to the high photo-meson optical depth at energies
$\epsilon_p>\epsilon_{p,\rm peak}$, we expect a neutrino spectrum similar
to the proton spectrum, $dn_\nu/d\epsilon_\nu\propto\epsilon_\nu^{-2}$,
above $\sim1$~TeV. The probability that a muon neutrino will
produce a high-energy muon in a terrestrial detector is \cite{Gaisser-rev}
$P_{\nu\mu}\approx1.3\times10^{-6}E_{\nu,\rm TeV}^{\beta}$, with $\beta=2$
for $E_{\nu,\rm TeV}<1$ and $\beta=1$ for $E_{\nu,\rm TeV}>1$. Thus,
for a flat neutrino spectrum above 1~TeV, the
muon flux at the detector is $\approx (P_0/E_0) {\cal F}_{\nu_\mu}$,
where $P_0/E_0=1.3\times10^{-6}{\rm TeV}^{-1}$. The number of events detected
in a burst of energy $E=10^{43}E_{43}$~erg is therefore
\begin{equation}
N_{\mu}\simeq 0.2\eta_{p,-1}\Gamma^{-1}\delta^3 D_{22}^{-2}E_{43} 
(A/1{\rm km}^2),
\label{Nmu}
\end{equation}
where $A$ is the effective detector area. For a persistent 
source (like SS433), the
corresponding rate of neutrino induced muon detection is
\begin{equation}
\dot N_{\mu}\simeq 0.2\eta_{p,-1}\Gamma^{-1}\delta^3 D_{22}^{-2}L_{j,38} 
(A/1{\rm km}^2){\rm day}^{-1}.
\label{eq:dotNmu}
\end{equation}

For sources directed along our sight line $\Gamma^{-1}\delta^3\sim 8\Gamma^2$ 
may exceed 100.  Thus, if the fraction 
$\eta_p$ exceeds a few percent, several neutrinos can be detected during
a typical outburst even from a source at 10 kpc. 
The typical angular resolution of the planned neutrino telescopes at TeV 
energies (e.g. \cite{halzen01}) should be $\theta\sim 1$ degree. 
The atmospheric neutrino background flux is 
$\Phi_{\nu,bkg} \sim 10^{-7}\epsilon_{nu,\rm TeV}^{-2.5}/{\rm cm^2 s\,sr}$, 
implying a number of detected background events 
$N_{bkg}\sim 3\times 10^{-2} (\theta /\hbox{deg})^2 t_{\rm day} {\rm km}^{-2}$
per angular resolution element over a burst duration 
$1t_{\rm day}$~day.  The neutrino signals above $\sim10$~TeV 
from a typical micro-quasar outburst should therefore be easily detected above 
the background. 

The duration of the neutrino burst should be of the order of the 
blob's ejection time.  It should precede the associated radio 
outburst, or the emergence of a new superluminal component, that 
originate from larger scales ($\sim 10^{15}$ cm), by several hours.  

As an example, let us consider the 1994 March 19 outburst 
observed in GRS 1915+105 \cite{MR94}.  This source is at a distance of
about 12 kpc, or $D_{22}=3.5$.  From the proper motions measured with the 
VLA, a speed of 0.92c and angle to the line of sight of 
$\theta\simeq70^{\circ}$ are inferred for the ejecta, corresponding to a 
Doppler factor $\delta\simeq0.58$.  A conservative estimate of the total 
energy released during this event yields $E_{43}=20$ \cite{LB96,MR99}.  From 
eq. (\ref{Nmu}) we then obtain, $N_{\mu}\simeq 0.03\eta_{p,-1}$ for a 1 km$^2$ 
detector.  At $\gtrsim$10 TeV this is above the background, but would require
an average of $\sim30$ outbursts for detection.  We note that if a similar 
event were to occur in a source that is located closer to Earth, at a 
distance of say 3 kpc as in the case of the superluminal source 
GRO J1655-40 (which has a similar Doppler factor), or from a jet oriented at 
a smaller angle, then a few or even a single outburst may produce several 
muon events in such a detector.
This suggests that even sources that are beamed away from 
us might be detectable during particularly strong outbursts. 

As a second example consider the source SS433. This source, at a distance of 
3 kpc, exhibits a steady jet that moves with a speed of $\sim0.3$c.  The 
presence of $H_{\alpha}$ lines indicates a baryonic content.  A conservative 
estimate of the jet's kinetic power yields $L_j\gtrsim10^{39}$ 
ergs s$^{-1}$ \cite{Ma84}.  Thus of order $10^3\eta_{p,-1}$ events per 
year per km$^2$ are anticipated. This rate is consistent with the MACRO upper limit 
\cite{MACRO}.  Emission of TeV neutrinos from this source has been proposed
earlier \cite{Eich80}, although the mechanism invoked in this ref. 
is pp collisions.

Positive detection of TeV neutrinos from microquasar jets would imply a 
baryonic content 
(althogh non-detection does necessarily imply a pair dominated jet), 
and would have important consequences for the mechanisms
responsible for the ejection and confinement of jets. It would also provide
important information concerning dissipation and particle acceleration 
in shocks.
Contemporaneous detections of gamma-rays may even enable us to infer the
relative efficiencies at which electrons and protons are accelerated in 
shocks by comparing the total energies emitted as 
neutrinos and gamma-rays during outbursts, thereby resolving a long 
standing issue.

We thank F. Aharonian, R. Blandford, C. Dermer, and D. Eichler for 
useful comments.
AL acknowledges support by the Israel Science Foundation.
EW is partially supported
by BSF Grant 9800343, AEC Grant 38/99 and MINERVA Grant.

\end{document}